\documentclass{Interspeech}

\interspeechcameraready

\title{InfiniteAudio: Infinite-Length Audio Generation with Consistency}

\author[affiliation={1}]{Chaeyoung}{Jung}
\author[affiliation={1}]{Hojoon}{Ki}
\author[affiliation={1}]{Ji-Hoon}{Kim}
\author[affiliation={1}]{Junmo}{Kim}
\author[affiliation={1}]{Joon Son}{Chung}

\affiliation{}{Korea Advanced Institute of Science and Technology, South Korea}

\email{\{codud9914, 
joon4366,
jh.kim, junmo.kim, joonson\}@kaist.ac.kr}
\keywords{text-to-audio generation, long generation, diffusion models}

\usepackage{comment}

\begin{document}

\maketitle

\begin{abstract}
    This paper presents InfiniteAudio, a simple yet effective strategy for generating infinite-length audio using diffusion-based text-to-audio methods. Current approaches face memory constraints because the output size increases with input length, making long duration generation challenging. A common workaround is to concatenate short audio segments, but this often leads to inconsistencies due to the lack of shared temporal context. To address this, InfiniteAudio integrates seamlessly into existing pipelines without additional training. It introduces two key techniques: FIFO sampling, a first-in, first-out inference strategy with fixed-size inputs, and curved denoising, which selectively prioritizes key diffusion steps for efficiency. Experiments show that InfiniteAudio achieves comparable or superior performance across all metrics. Audio samples are available on our project page\footnote{https://mm.kaist.ac.kr/projects/InfiniteAudio/}.
   
\end{abstract}

\section{Introduction}
Diffusion models~\cite{ho2020denoising, song2020score} have gained significant attention for their ability to generate high-quality and diverse outputs, achieving state-of-the-art performance across various domains, including image~\cite{dhariwal2021diffusion, rombach2022high, saharia2022photorealistic, nichol2021glide}, video~\cite{ho2022video, singer2022make, wang2023modelscope,yang2023diffusion,wang2024lavie,bar2024lumiere,chen2023videocrafter1}, and audio~\cite{popov2021grad, kim2022guided, li2024styletts, jeong2021diff,liu2023diffvoice, lee2024seeing, jung2024flowavse}.
However, their high computational cost limits their practicality in many applications. To address this, the Latent Diffusion Model (LDM)~\cite{rombach2022high} was introduced, utilizing a compressed latent space to improve efficiency. This approach significantly reduces computational overhead while preserving high-fidelity outputs, making it more practical for image and video generation~\cite{blattmann2023align, podell2023sdxl, zhou2022magicvideo}.

Beyond image and video generation, LDMs have also become a cornerstone in audio synthesis, particularly in text-to-audio (TTA) generation~\cite{liu2023audioldm, huang2023make, lee2024voiceldm, liu2024audioldm, ghosal2023text, yang2023diffsound,yuan2024retrieval, kreuk2022audiogen,liu2024audiolcm,jung2024voicedit}, which produces realistic audio from textual prompts.
By leveraging LDM’s efficiency and generative power, TTA models have advanced significantly, incorporating Contrastive Language-Audio Pretraining (CLAP)~\cite{wu2023large} to enhance alignment between textual descriptions and generated audio~\cite{liu2023audioldm, huang2023make, yuan2024retrieval}.
Additionally, large language models (LLMs) have been integrated into TTA frameworks~\cite{liu2024audioldm, ghosal2023text}, improving text embeddings and enabling more precise interpretation of complex prompts. This advancement allows TTA models to generate audio that more accurately reflects the intended context.

Despite significant advancements in TTA generation, existing models based on diffusion approaches face substantial challenges in generating long-duration audio~\cite{liu2023audioldm, lee2024voiceldm, kreuk2022audiogen, guo2024audio}. The core issue arises from the inherent design of diffusion models, which require the input and output dimensions to remain identical throughout the process. As a result, generating longer audio necessitates a proportional increase in input size, leading to memory constraints.
A common workaround involves concatenating short audio clips produced by existing TTA models to create longer sequences. However, this approach often suffers from temporal inconsistencies between segments, resulting in unnatural and discontinuous audio streams. The lack of shared temporal context across clips makes it difficult to maintain coherence and smooth transitions, further limiting the practical application of these models for long-form audio generation.

\begin{figure*}[!t]
  \centering
\includegraphics[width=0.9\linewidth]{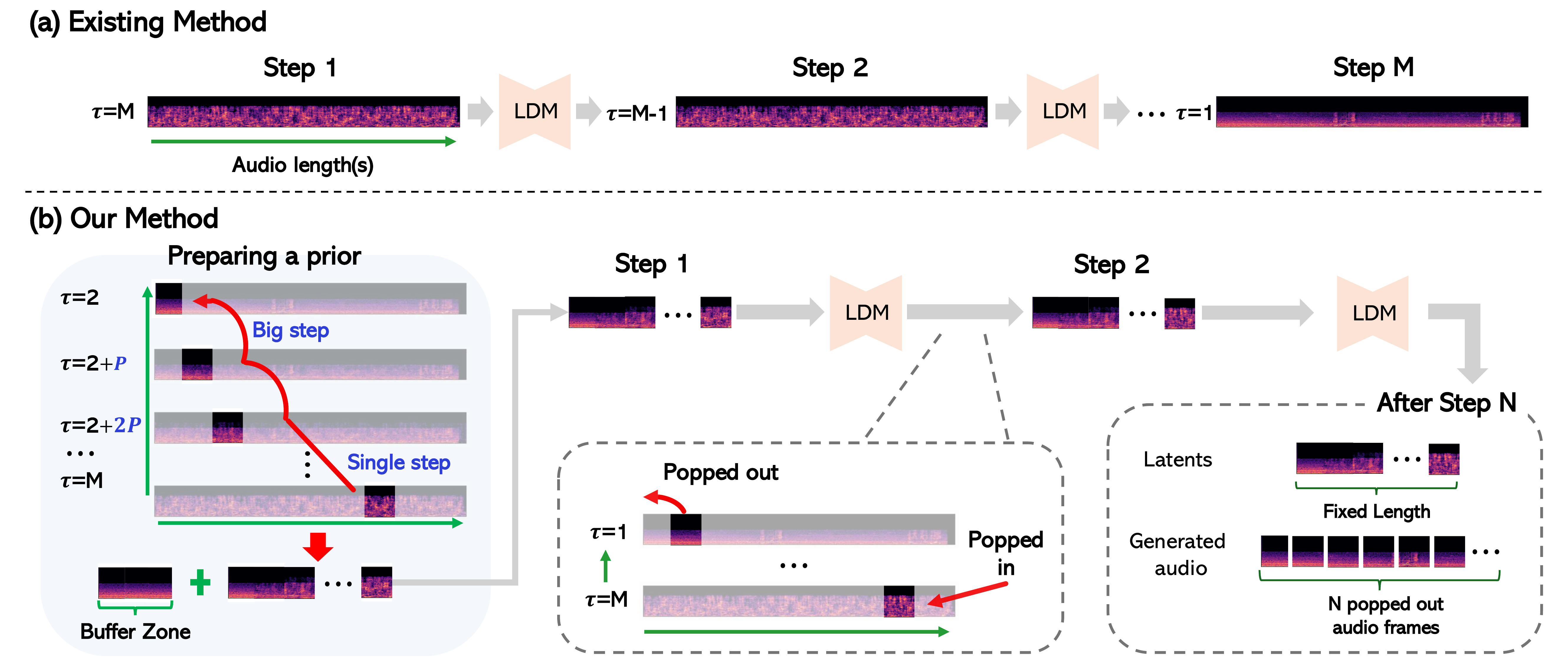}
\vspace{-2mm}  \caption{Comparison of Existing Methods and InfiniteAudio. Traditional methods apply uniform diffusion timesteps across all input latents, whereas InfiniteAudio dynamically selects timesteps based on their importance. This adaptive approach enables the generation of theoretically infinite audio while maintaining a fixed input size, ensuring both efficiency and high-quality synthesis.}
  \vspace{-2mm}
  \label{Fig1}
\end{figure*}

To address these limitations, we propose InfiniteAudio, a novel inference technique designed to generate long and temporally consistent audio. InfiniteAudio overcomes the memory constraints of diffusion models by employing a first-in-first-out (FIFO) mechanism with a fixed input size. This approach incrementally adds noise to parts of the existing model's predictions, as illustrated in Fig.~\ref{Fig1}. Unlike traditional diffusion models that begin with a uniform noise prior, InfiniteAudio uses priors with varying noise levels.
During inference, the fully denoised portion of the input at the start is discarded, and new latent noise is appended at the end. This progressive replacement of input data enables InfiniteAudio to generate arbitrarily long audio sequences while maintaining temporal consistency and a constant memory footprint. By rethinking the inference process, InfiniteAudio eliminates the reliance on concatenating short audio clips and enables the direct generation of long-duration audio. This approach not only resolves the temporal inconsistencies of existing methods but also offers a solution for generating coherent, seamless audio over extended durations.

Although FIFO-Diffusion~\cite{kim2024fifo}, a method developed for text-to-video (TTV) generation, employs a FIFO generation mechanism, it still utilizes all diffusion sampling steps during inference. In contrast, as illustrated in Fig.~\ref{Fig1}, our approach introduces curved denoising, a novel technique that selectively prioritizes the most critical diffusion steps identified through self-attention maps, rather than uniformly applying all steps. This targeted strategy preserves high-quality generation while substantially reducing the number of sampling steps, leading to a more efficient inference process.

Our contributions are summarized as follows. First, we first propose InfiniteAudio, a method capable of generating long-duration audio sequences without requiring additional training. It effectively addresses the memory limitations inherent in existing diffusion-based TTA models. 
Second, we introduce curved denoising, a selective sampling technique that focuses on critical diffusion steps instead of utilizing all steps, as in FIFO-Diffusion~\cite{kim2024fifo}, resulting in improved sampling efficiency. 
Lastly, our method can be seamlessly integrated into existing text-to-audio generation baselines.

\section{Method}

\subsection{Preliminaries}

We provide a comprehensive overview of existing TTA generation models, which synthesize realistic audio from text prompts $y$ by representing audio as a 2D mel-spectrogram, capturing both time and frequency dimensions.

Most TTA models share a common architecture, consisting of audio $f_{audio}(\cdot)$ and text encoders $f_{text}(\cdot)$, a LDM, an audio decoder, and a vocoder. 
The encoders map text and audio inputs into a latent space, where the LDM is trained to iteratively refine a perturbed latent representation $\mathbf{z}_{\tau}$. Here, $\tau \sim \mathcal{U}([1, ..., M])$ represents the diffusion timestep, controlling the level of noise at each step.  
The LDM progressively denoises $\mathbf{z}_{\tau}$ from a noisy state back to a clean latent representation $\mathbf{z}_1$. 
Once $\mathbf{z}_1$ is obtained, the audio decoder reconstructs the mel-spectrogram $a\in \mathbb{R}^{T\times F}$ from $\mathbf{z}_{1} \in \mathbb{R}^{C\times \frac{T}{r} \times \frac{F}{r}}$, where $T$ and $F$ denote the time and frequency dimensions, $C$ is the number of channels, and $r$ is the compression factor. Finally, the vocoder converts the reconstructed mel-spectrogram into a waveform, producing the final audio output.

To train the model, Gaussian noise is gradually added to the latent representation over multiple timesteps. The model then learns to iteratively remove the noise to reconstruct the original clean representation.
Given a random noise sample $\bm{\epsilon} \sim \mathcal{N}(\mathbf{0},\mathbf{I})$, where $\mathcal{N}(\mathbf{0},\mathbf{I})$ denotes a standard normal distribution and a text condition $\mathbf{c} = f_{text}(y)$ obtained from the text encoder, the model is trained to minimize the following denoising loss function: 
\begin{equation}
\mathcal{L} = \mathbb{E}_{\mathbf{z}_0, \bm{\epsilon}, \tau} \left[ | \bm{\epsilon} - \bm{\epsilon}_\theta(\mathbf{z}_{\tau}, \tau, \mathbf{c}) |^2_2  \right],
\end{equation}
where $\bm{\epsilon}_\theta$ represents the model’s predicted noise at timestep $\tau$.

\subsection{InfiniteAudio}
\label{InfiniteAudio}
In this section, we introduce InfiniteAudio, a technique for generating long-duration audio while maintaining a fixed memory footprint. Our approach leverages pre-trained diffusion-based TTA models and operates without requiring additional training. We focus on two key models, AudioLDM and VoiceLDM, and demonstrate how our inference strategy effectively mitigates their inherent memory constraints.

\subsubsection{FIFO sampling}
\label{diffusiontimestep}
Generating long audio sequences with diffusion models is challenging due to their high memory requirements. Recently, FIFO-Diffusion~\cite{kim2024fifo} has addressed this issue in video generation by utilizing a fixed-size input, where each frame is assigned a different diffusion timestep. This allows the model to apply multiple diffusion steps simultaneously across the input frames.
This approach enables the generation of theoretically infinite video by applying a FIFO sampling strategy. 

We adapt this concept to audio generation by initiating the diffusion process with a fixed-length audio segment, where each segment is treated similarly to video frames. The input latent  $\mathbf{z}_\tau \in \mathbb{R}^{C \times \frac{T}{r} \times \frac{F}{r}}$ is treated as $\frac{T}{r}$ audio frames, analogous to video frames. Here, each compressed mel-spectrogram frame corresponds to $\mathbf{z}_1^i \in \mathbb{R}^{C \times 1 \times \frac{F}{r}}$, where $i \in [1, \frac{T}{r}]$. This structure enables us to apply multiple diffusion steps across the audio segment in a similar manner to video generation.

For infinite audio generation, noise is progressively added to the input audio frames over time, except for the initial frames, which act as a "buffer zone" and are not perturbed by noise. Since no additional training occurs in InfiniteAudio, using different diffusion timesteps during inference can introduce a performance gap, as shown in~\cite{kim2024fifo}. The buffer zone helps mitigate this by ensuring that the same timesteps used during training are applied to the initial frames, reducing inconsistencies.

Beyond the buffer zone, the earlier frames are nearly fully predicted and the later frames are treated as Gaussian noise. At the inference stage, the input consists of the buffer frames and frames with increasing noise levels.
As represented in Fig.~\ref{Fig1} (b), after each inference step, the first frame following the buffer zone reaches diffusion timestep $\tau =1$ and is then removed. To maintain the same input size, we insert a new noisy frame at the last position. By iteratively repeating this process, we generate $N$ frames in $N$ inference steps, enabling seamless and consistent infinite audio generation through continuous synthesis.

\subsubsection{Curved Denoising with Reduced Sampling Steps}
To address memory limitations, InfiniteAudio maintains a constant input size during inference, independent of the output length. However, using the full set of diffusion timesteps still requires long input sequences, as more audio segments must be generated by existing TTA models.
To overcome this, InfiniteAudio prioritizes critical diffusion step regions while reducing emphasis on less important ones. By leveraging deterministic denoising~\cite{song2020denoising}, existing models can perform inference efficiently, skipping unnecessary steps while maintaining high-quality output. Similarly, we eliminate non-essential steps while preserving sample fidelity, guided by self-attention maps that highlight key regions for generation.

\begin{figure}
  \centering
\vspace{-3mm}\includegraphics[width=1.0\linewidth]{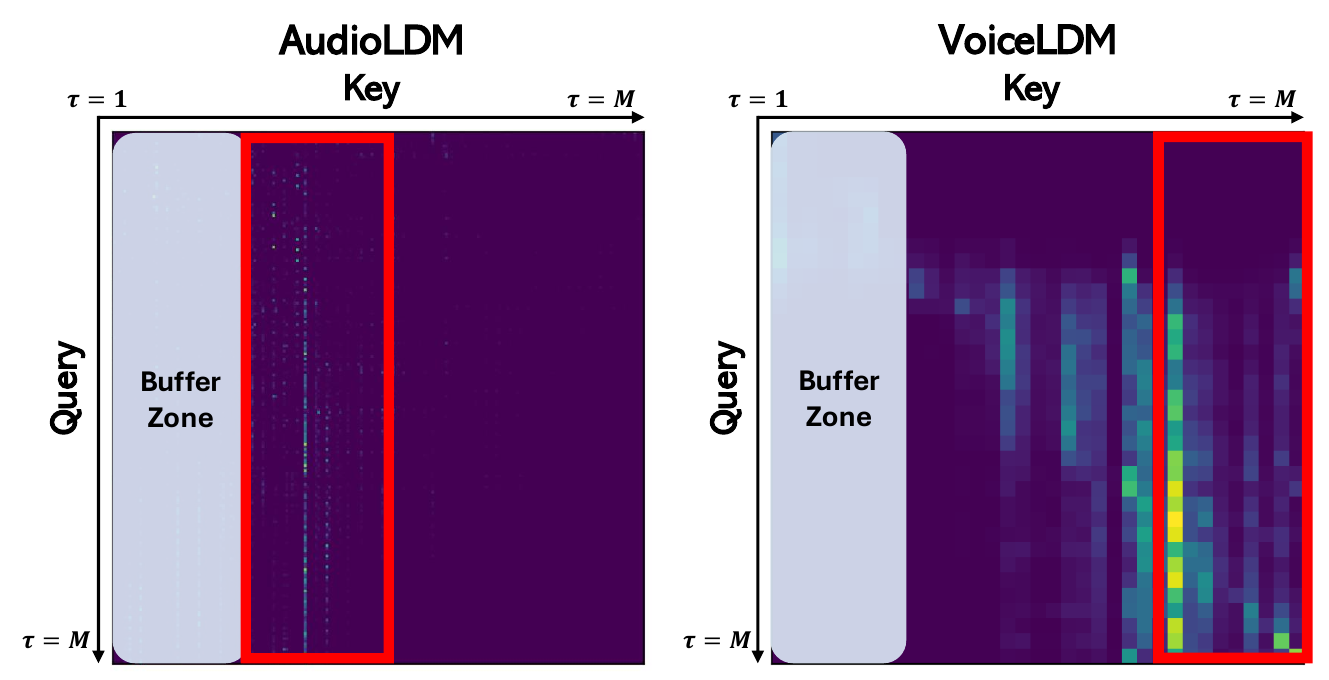}
\vspace{-5mm}
  \caption{Attention maps indicating the importance of timesteps in input sequences. In AudioLDM, the query primarily attends to the initial portions of the key. In contrast, VoiceLDM exhibits a stronger correlation with the later key segments to its query, highlighting a distinct attention distribution pattern.}
\vspace{-5mm}
  \label{attentionmap}
\end{figure}  
We partition the diffusion sampling steps into three distinct regions based on the original timestep distribution: initial, middle, and final. The initial region corresponds to the early stages of sampling, where diffusion timesteps are close to $\tau$, while the final region represents the later stages, where timesteps approach $1$.
To identify critical sampling regions during inference for AudioLDM and VoiceLDM, we analyze self-attention map scores within the InfiniteAudio framework. In the self-attention mechanism, attention scores quantify the relevance between a query and a key vector, determining the influence of one element on another within the sequence. By examining self-attention maps in U-Net decoder modules throughout the diffusion process, we can pinpoint key frames that exert the most influence on query frames in each model.

As shown in Fig.~\ref{attentionmap}, model behavior varies significantly based on configuration. In AudioLDM, query sequences are primarily influenced by initial key sequences, corresponding to earlier frames with diffusion timesteps close to $1$. In contrast, VoiceLDM exhibits greater sensitivity to later key sequences, which represent noisier inputs with timesteps approaching $M$. Additionally, some initial frames fall within a transitional buffer zone, so our analysis focuses on regions beyond this buffer for greater clarity.
Based on these observations, we compute the average attention scores across the initial, middle, and final regions, as illustrated in Fig.~\ref{attentionmap}. We then involve timesteps to regions with higher average attention scores, while skipping less critical regions using a factor of $P$. This adaptive timestep allocation effectively reduces both the number of inference steps and input size to about 3 seconds. Our curved denoising method preserves output quality while requiring fewer computations, enhancing overall efficiency.

\section{Experiment}

\subsection{Experimental Settings}
\subsubsection{Datasets and Baselines.}
To evaluate our method on TTA generation, we utilize 500 audio-text pairs from the 975 test files in the Audiocaps dataset~\cite{kim-NAACL-HLT-2019}, which is commonly used for assessing TTA models. Given that our method relies heavily on the performance of existing baselines, we exclude the bottom 20 percent of audio-text pairs with the lowest CLAP scores, as predicted by AudioLDM and VoiceLDM. From the remaining pairs, 500 are randomly selected for evaluation.
For comparison, we assess the performance of InfiniteAudio against two publicly available TTA models: AudioLDM\footnote{https://github.com/haoheliu/AudioLDM} and VoiceLDM\footnote{https://github.com/glory20h/VoiceLDM}.

\subsubsection{Evaluation Metrics.}
We evaluate audio quality and text-audio alignment using standard quantitative metrics, including Frechet Distance (FD), Kullback-Leibler (KL) divergence, and the CLAP score~\cite{liu2023audioldm, lee2024voiceldm, vyas2023audiobox}. Frechet Distance and Kullback-Leibler divergence quantify how closely the generated audio matches the ground truth, where lower values indicate better performance. The CLAP score measures the relevance of the generated audio to the text prompt, with higher values being preferable.
For subjective evaluation, we assess overall quality (OVL) and relevance to the input text (REL). Both metrics were rated on a scale of 1 to 5 by 20 domain experts using 30 speech samples.

\begin{table*}
\centering
\scriptsize
\vspace{-3mm}
\def\arraystretch{1.14}% 
\caption{Quantitative evaluation of TTA.
Our method demonstrates performance comparable to both models, even surpassing the original inference results. Additionally, the use of equally spaced timesteps, as suggested in~\cite{kim2024fifo}, is considered.}
\vspace{-3mm}
\begin{tabular}{l|ccc|cc}

\toprule
Method  & CLAP$\uparrow$   & FD$\downarrow$ & {KL$\downarrow$} & OVL$\uparrow$ &REL$\uparrow$ \\ \midrule

Ground Truth  & 0.5276 & NA    & {NA} &4.11$\pm$0.22 &4.03$\pm$0.25  \\ \midrule

AudioLDM~\cite{liu2023audioldm}  & \textbf{0.4908}       &  \underline{44.6689}   & {\underline{2.0805}} &{\bf 3.03$\pm$0.23 } &{\bf 3.06$\pm$0.21}  \\ 
 w/ Equally spaced timesteps~\cite{kim2024fifo} & 0.3832 & 54.7479    & {2.4013}  &2.19$\pm$0.21 &2.33$\pm$0.23  \\ 
 w/ Middle focused timesteps   & 0.3979 & 56.7792    & {2.6077}  &2.06$\pm$0.19 &2.18$\pm$0.20 \\ 
\textbf{ w/ Last focused timesteps (Ours)}   & \underline{0.4559} & \textbf{43.3788}    & {\textbf{1.9650}} &\underline{2.63$\pm$0.18} &\underline{2.80$\pm$0.21}  \\
 w/ Initial focused timesteps  & 0.3110 &67.0704     & {2.9838} &2.13$\pm$0.19 &2.07$\pm$0.20  \\ 
 \midrule
VoiceLDM~\cite{lee2024voiceldm}  & \textbf{0.4199} & \textbf{51.4019}    & {\textbf{2.2749}}  &{\bf 2.53$\pm$0.24}  &\underline{2.41$\pm$0.21} \\ 
 w/ Equally spaced timesteps~\cite{kim2024fifo}    & 0.3729 & 59.1521    & {2.4477} &2.20$\pm$0.21 &2.33$\pm$0.22  \\ 
 w/ Middle focused timesteps  & 0.3779 & 56.7321    & {2.4622} &2.10$\pm$0.20 &\underline{2.41$\pm$0.22}  \\ 
 w/ Last focused timesteps   & 0.3542 & 64.8813    & {2.6227}   &\underline{2.38$\pm$0.23} &2.24$\pm$0.21 \\ 
\textbf{ w/ Initial focused timesteps (Ours)}   & \underline{0.4107} &\underline{51.5047} & {\underline{2.3498}}  &\underline{2.38$\pm$0.23} &{\bf 2.48$\pm$0.21} \\ 
\bottomrule
\end{tabular}
\vspace{-2mm}
\label{table1_TTA}
\end{table*}

\subsection{Quantitative Results}

\begin{figure}
  \centering
\vspace{-5mm}\includegraphics[width=0.95\linewidth]{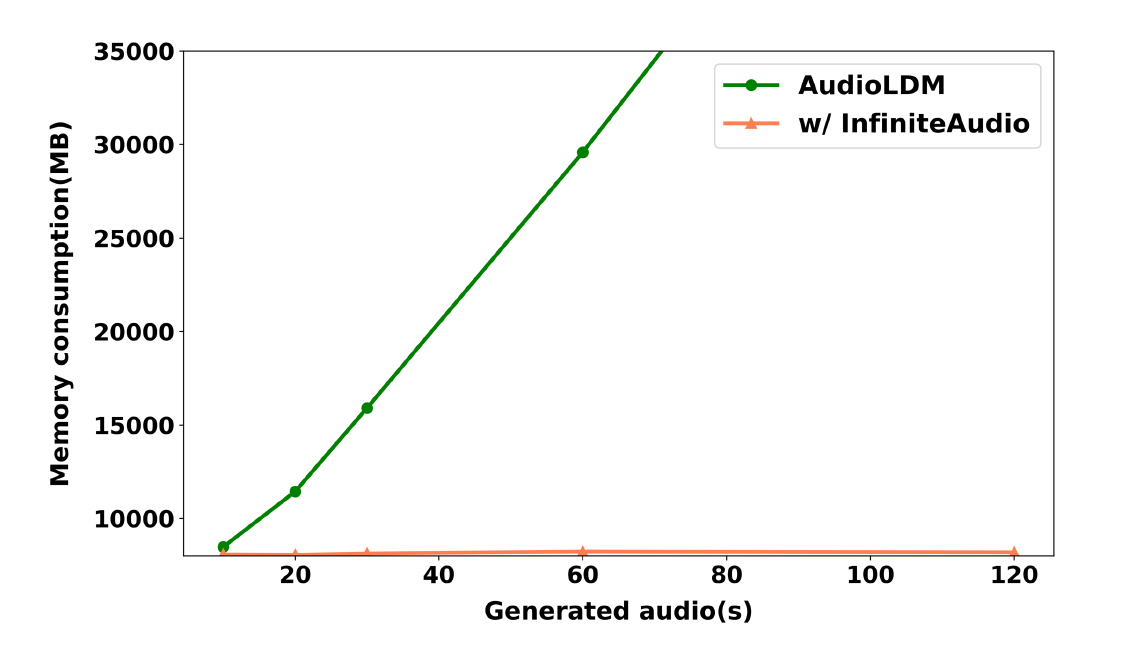}
  \vspace{-5mm}
 \caption{Memory consumption comparison between AudioLDM~\cite{liu2023audioldm} and our method}
 \vspace{-5mm}
  \label{mem}
\end{figure}

\subsubsection{Memory Consumptions.}
We compare the memory consumption of AudioLDM with our method. 
AudioLDM's memory usage grows with the length of the generated audio, while our method maintains constant memory usage regardless of audio length, as shown in Fig.~\ref{mem}.

\subsubsection{Evaluation of Curved Denoising.}
We evaluate the effectiveness of our curved denoising strategy in Tab.~\ref{table1_TTA}, comparing 10-second audio samples generated using different sampling methods. 
Despite requiring no additional training, our method not only matches the performance of existing models but also achieves higher scores. By incorporating self-attention relevance, our approach outperforms methods that use equally spaced timesteps, such as FIFO-Diffusion~\cite{kim2024fifo}, as well as other strategies with the same number of steps. 

\subsection{Qualitative Results}
\subsubsection{Sampling Strategies.}
\begin{figure}
\centering
\vspace{-3mm}
\includegraphics[width=1.0\linewidth]{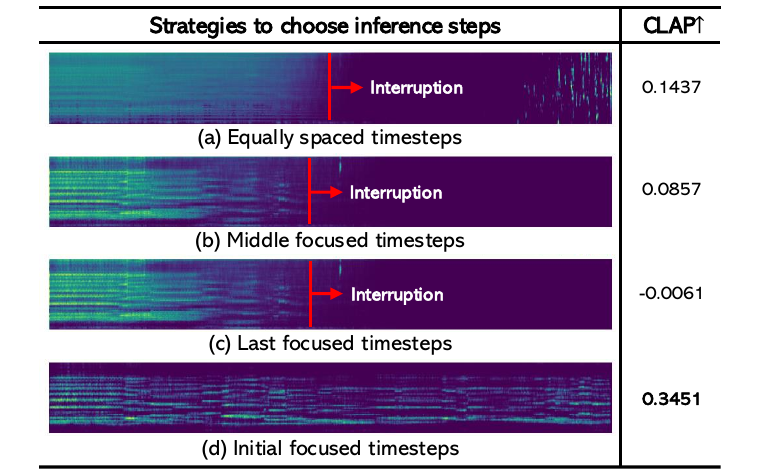}
\vspace{-5mm}
\caption{Analysis of different diffusion sampling strategies for VoiceLDM~\cite{lee2024voiceldm}.} 
\vspace{-3mm}

\label{Qualitative1}
\end{figure}

As shown in Fig.~\ref{Qualitative1}, unlike other strategies that exhibit interruptions in the generated audio, as seen in the spectrograms, our method ensures seamless audio generation. This is supported by both the spectrogram analysis and the improved CLAP score.

\begin{table}
\centering
\scriptsize
\caption{Comparison of sampling steps in VoiceLDM~\cite{lee2024voiceldm}.
InfiniteAudio achieves superior results for 10 second audio generation while requiring fewer than 150 sampling steps, demonstrating enhanced efficiency.}
\vspace{-3mm}
\begin{tabular}{l|ccc}
\toprule
Sampling steps & CLAP$\uparrow$ & FD$\downarrow$ & KL$\downarrow$ \\ \midrule
w/ 200 equally spaced steps            &  0.3923   & {53.0555}   & \textbf{2.3334}   \\
w/ 250 equally spaced steps            &  \underline{0.3941}    & \textbf{50.5447}   &    2.3937 \\   \midrule

\textbf{InfiniteAudio}  &  \textbf{0.4107}    &\underline{51.5047}    & \underline{2.3498}   \\ 
\bottomrule

\end{tabular}
\label{wrap-tab:5}
\vspace{-4mm}
\end{table}
\begin{figure}
\centering
\vspace{-3mm}
\includegraphics[width=1.0\linewidth]{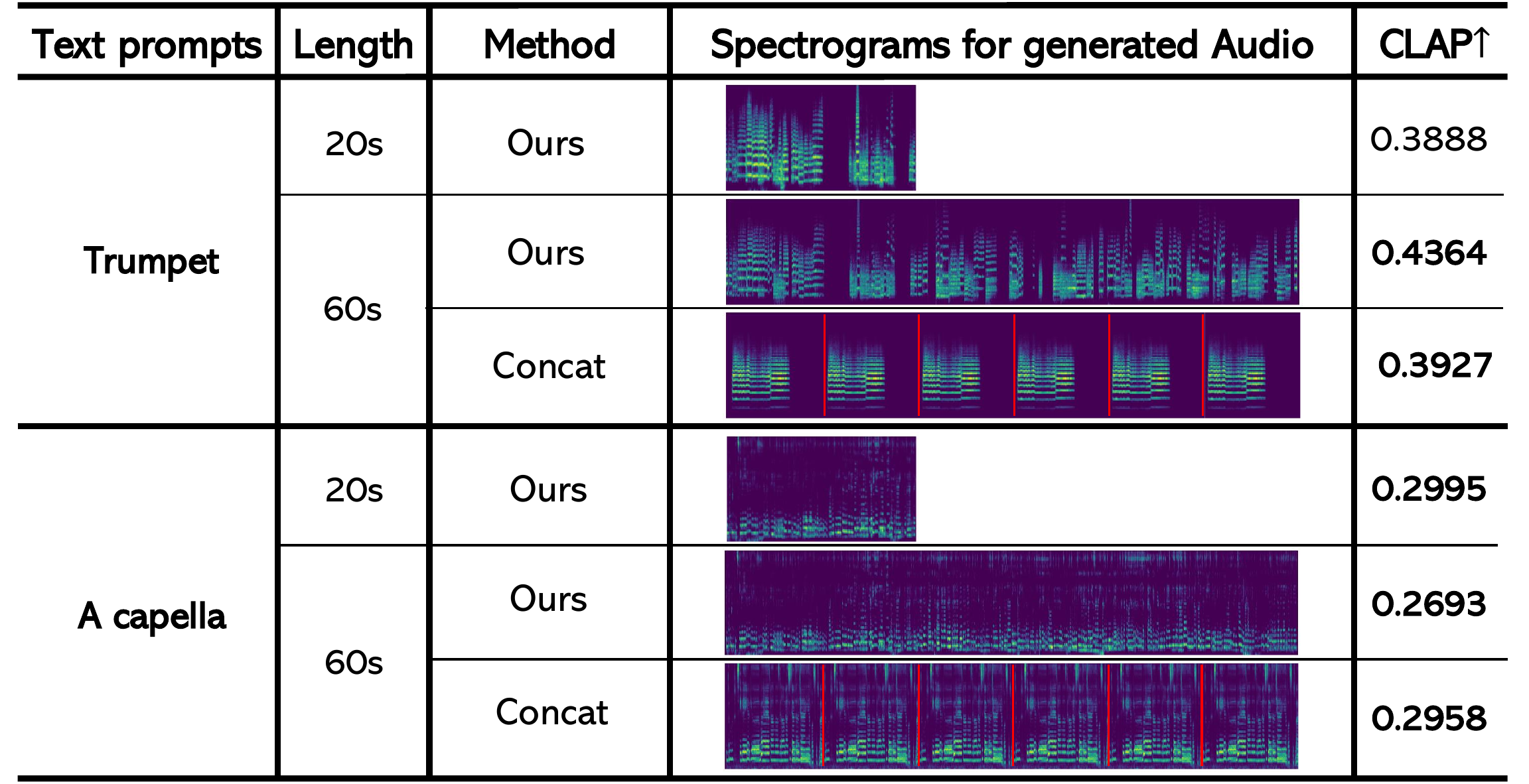}
\vspace{-6mm}
\caption{Comparison of audio generated by AudioLDM~\cite{liu2023audioldm} using InfiniteAudio and the concatenation method, demonstrating InfiniteAudio’s superior long-duration generation.}
\label{Quantitative}
\vspace{-1mm}
\end{figure}

\subsubsection{Long Generation Quality.}
Fig.~\ref{Quantitative} presents the mel spectrograms of audio generated by AudioLDM~\cite{liu2023audioldm} at different lengths using InfiniteAudio, compared to audio produced via the concatenation method. The results demonstrate that InfiniteAudio generates high-quality, long-duration audio while preserving stable CLAP scores across varying lengths. Moreover, it maintains seamless consistency and natural coherence throughout the entire audio, whereas the concatenation method results in repetitive patterns and temporal inconsistencies, as highlighted by the red line.

\subsection{Analysis on Sampling Steps and Audio Length}
InfiniteAudio is designed to optimize sampling efficiency by minimizing the number of sampling steps while maintaining high audio quality. Unlike methods that extend sampling to 200 or 250 steps with equally spaced timesteps, our approach achieves consistently solid performance across all metrics while using fewer than 150 steps, as shown in Tab.~\ref{wrap-tab:5}.

InfiniteAudio also excels in generating longer audio sequences without sacrificing quality. As shown in Tab.~\ref{wrap-tab:6}, it delivers results comparable to the fixed 10-second generation across varying lengths, ranging from 10 to 20 seconds. Notably, the CLAP score for this experiment is calculated using a different checkpoint than the one used in other tables, as evaluating varying audio lengths requires a distinct CLAP model\footnote{https://github.com/LAION-AI/CLAP}.

\begin{table}
\centering
\scriptsize
\vspace{-1mm}
\caption{Comparison of generated audio lengths between a fixed 10 second and variable-length in AudioLDM~\cite{liu2023audioldm}.}
\vspace{-3mm}
\begin{tabular}{l|ccc}
\toprule
Generated audio length & CLAP$\uparrow$ & FD$\downarrow$ & KL$\downarrow$ \\ \midrule
Fix          &  0.3207   & \textbf{43.3788}   & {1.9650}   \\
Various         &  \textbf{0.3259}    & {44.3701}   &    \textbf{1.9044}\\
\bottomrule
\end{tabular}
\vspace{-5mm}
\label{wrap-tab:6}
\end{table}
\section{Conclusion}
We introduce InfiniteAudio, a novel inference method for generating infinitely long, consistent audio using pretrained text-to-audio models. By maintaining a fixed memory footprint, InfiniteAudio overcomes memory constraints in existing models and integrates seamlessly with diffusion-based TTA approaches. Despite relying solely on inference techniques, it achieves superior performance, opening new possibilities for continuous and coherent long-form audio generation.

\section{Acknowledgements}
This work was supported by the National Research
Foundation of Korea (NRF) grant funded by the Korea
government (MSIT) (No. RS-2023-00212845, Multimodal
Speech Processing for Human-Computer Interaction).
%\ifinterspeechfinal
%     The Interspeech 2025 organisers
%\else
%     The authors
%\fi
%would like to thank ISCA and the organising committees of past Interspeech conferences for their help and for kindly providing the previous version of this template.

\bibliographystyle{IEEEtran}
\bibliography{mybib}

\end{document}